# Multiferroic metal—PbNb$_{0.12}$Ti$_{0.88}$O$_{3-\delta}$ films on Nb-doped STO

*Hongbao Yao[†, ‡], Jiesu Wang[†], Kuijuan Jin[\*, †, ‡, §], Qinghua Zhang[†], Wenning Ren[†, ‡], V. Pazhanivelu[†], Lin Gu[†, ‡], Chen Ge[†, ‡], ErJia Guo[†, ‡], Xiulai Xu[†, ‡, §], Can Wang[†, ‡, §], and Guozhen Yang[†, ‡]*

[†] Institute of Physics, Chinese Academy of Sciences, Beijing 100190, China

[‡] University of Chinese Academy of Sciences, Beijing 100049, China

[§] Songshan Lake Materials Laboratory, Dongguan, Guangdong 523808, China

ABSTRACT: Ferroelectricity—the switchable intrinsic electric polarization—has not yet been attained in a metal experimentally and is in fact generally deemed irreconcilable with free carriers, although polar metal has been achieved recently. Multiferroic metal has never even been proposed though multiferroics have been widely investigated. Here we report a room-temperature coexistence of multiferroicity and metallic behavior in PbNb$_{0.12}$Ti$_{0.88}$O$_{3-\delta}$ films. The oxygen-vacancy-induced electrons become delocalized and ameliorate the ferromagnetic properties of these films, whereas they fail to vanish the polar displacements nor the individual dipole in each unit cell. This concurrent appearance of multiferroicity and metallicity is also confirmed by our first-principles calculation performed on 12.5% Nb-doped PbTiO$_3$ with oxygen vacancies. These findings break a path to multiferroic metallic materials and offer a potential application for multiferroic spintronic devices.

## 1. INTRODUCTION

Ferroelectric-like structure in a metal, proposed by Anderson and Blount in 1965[1], had been thought mutually incompatible because the free carriers would extinguish the polarized charges[2]. Surprisingly, several polar metals have been accomplished in recent years. For example, the ferroelectric-like structural transition in high-pressure-synthesized metallic LiOsO$_3$[3], the retained ferroelectricity in a high electron concentration titanium oxide BaTiO$_{3-\delta}$[4], the ferroelectric switching in two- or three-layered metallic WTe$_2$[5], and the simultaneous coexistence of a polar structure and metallicity in BaTiO$_3$/SrTiO$_3$/LaTiO$_3$[6] superlattice, NdNiO$_3$[7], and Ca$_3$Ru$_2$O$_7$[8] were reported. Gu et al.[9] have proved the PbTi$_{0.88}$Nb$_{0.12}$O$_3$ ceramic target with 12% Nb for deposition was the optimized one to show the coexistence of polar and metallicity along the out-of-plane (OP) direction. In addition to experimental realizations, a polar metal structure has also aroused broad interest in theoretical design for their potentially unconventional optical responses, magnetoelectricity, and superconductivity properties[10-16].

On the other hand, multiferroic materials—possessing more than one of the ferroelectricity, magnetic ordering, and ferroelasticity simultaneously—have been found in heterostructures and composites[17-23], especially in that with perovskite structures, like YMnO$_3$[24], BiFeO$_3$[25,26], BiMnO$_3$[27-30], and EuTiO$_3$[31]. These compounds bearing rich functionality present promising applications in sensors and high-performance storage devices. Thus, the achievement of multiferroic metal promotes to reveal the mechanism of ferroelectric, magnetic, and structural order parameters coupling, and offer new design to spintronic multiferroic devices.

As far as we know, multiferroic metal has never even been proposed yet. Here, we report an observation at room-temperature of the coexistence for the multiferroicity and metallicity in PbNb$_{0.12}$Ti$_{0.88}$O$_{3-\delta}$ (PNTO$_{3-\delta}$) films on Nb-doped SrTiO$_3$ (SNTO) substrates. Distinguishable magnetic hysteresis loops are captured even at the ambient temperature. Piezoresponse force microscopy (PFM) results prove the intrinsic switchable electric polarizations in these films, together with the characterization by scanning transmission electron microscopy (STEM) and second-harmonic generation (SHG) technology. Distinct ferroelectric hysteresis loops demonstrate the ferroelectricity with spontaneous polarization. Moreover, electric transport results reveal evident metallic behaviors for these PNTO$_{3-\delta}$ films in the direction perpendicular to the sample surface. Our first-principles calculation further illustrates that 12.5% Nb-doped PbTiO$_3$ with oxygen vacancies (V$_o$) is a half metal with spontaneous polar displacements, and that the delocalized electrons at the $d_{xy}$ orbitals of Nb and Ti atoms induce the ferromagnetism.

## 2. EXPERIMENTAL SECTION

**2.1. Film Growth.** The PNTO$_{3-\delta}$ films were deposited on (001)-oriented 0.7 *wt*% SNTO substrates under 6 Pa and 4 Pa oxygen partial pressures using a Laser-MBE system. The deposition temperature was set to 520 °C and the laser energy density was fixed at ~1.2 J/cm$^2$. Before cooling down to room temperature, all heterostructures were annealed in situ at the deposition condition for 20 min to ensure the stoichiometry. The thicknesses of these films are ~100 nm, controlled by the deposition time.

**2.2. X-ray Characterization.** The X-ray diffraction (XRD) and X-ray diffractometry reciprocal space mapping (RSM) analyses were performed using a Rigaku SmartLab (8 kW) High-resolution (Ge 220 × 2) X-ray Diffractometer, with the wavelength of the X-ray is 0.154 nm.

**2.3. Physical Properties.** Electric transport properties were measured using a Physical Properties Measurement System (PPMS, Quantum Design) with the temperature ranging from 10 K to 360 K. Au/ PNTO$_{3-\delta}$/SNTO sandwich structures were made with the Au electrode diameter of ~750 um. During the experiments, a current of 1 μA was applied using a SourceMeter (2400, Keithley) and the voltages were measured by a NanovoltMeter (2182A, Keithley). PPMS was also used to acquire the magnetic properties and the diamagnetic contributions from the substrates were removed.

**2.4. Scanning Transmission Electron Microscopy.** Specimens for STEM were gained by mechanically polishing the heterostructures to about 20 μm. Central parts of the specimens were further reduced by precision argon-ion milling, until transparent for electron beams. The atomic structures of these heterostructures were characterized using an ARM-200CF (JEOL, Tokyo, Japan) transmission electron microscope operated at 200 keV and equipped with double spherical aberration (Cs) correctors. The collection angle of high-angle annular dark-field (HAADF) images was 90-370 mrad. All HAADF images were filtered using the HREM-Filters Pro/Lite released by HREM Research Inc. Atomic positions were determined by fitting with Moment Method & Contour using the CalAtom Software developed by Prof. Fang Lin.

**2.5. Piezoresponse Force Microscopy.** The PFM phase images were attained on a commercial atomic force microscope (AFM, Asylum Research MFP-3D). Ti/Ir-coated Si cantilevers (Olympus Electrilever) were used to collect and record the PFM images, which have the nominal spring constant of ~2 N/m and the tested free air resonance frequency of ~73 kHz.

**2.6. Ferroelectric Hysteresis Loops.** We employed a ferroelectric tester (Radiant Technologies, Premier II) to acquire the ferroelectric hysteresis loops of the Au/PNTO$_{3-\delta}$/SNTO sandwich structures under the capacitance configuration. The ferroelectric hysteresis loops were derived by subtracting one hysteresis loop from another inverse one to minimize the removable charges contribution on the polarization signals. The measuring frequencies were set to 10 kHz.

**2.7. Second-harmonic Generation.** Far-field SHG polarimetry measurements were performed in the typical reflection geometry[32]. The incident laser beam was generated by a Spectra Physics Maitai SP Ti:Sapphire oscillator with the central wavelength at 800 nm (~120 fs, 82 MHz). The incident laser powers were attenuated to 70 mW before being focused on the films. Both incidence angle and reflection angle were fixed at 45°. The polarization direction $\varphi$ of the incident light field was rotated by a $\lambda/2$ wave plate driven by a rotational motor. Generated second-harmonic light field reflected were firstly decomposed into *p*-polarized and *s*-polarized components, noted as *p*-out and *s*-out, respectively. After spectrally filtered, the second-harmonic signals were detected by a photo-multiplier tube. The read-out sums of frequency-doubled photons are proportional to the SHG responses. The polar plots were acquired through rotating $\varphi$ for *p*-out and *s*-out configurations. Specific SHG configuration and the fitting equations of the results can be found in our former researches[32].

## 3. THEORETICAL CALCULATIONS

The first-principles calculation was carried out using density functional theory (DFT) with Vienna ab initio simulation package (VASP). For simplicity, a doping concentration of 12.5% was adopted by constructing a 2 × 2 × 2 supercell of $PbTiO_3$. A Ti atom was replaced by Nb and one of the oxygen atoms in the structure was removed to simulate the oxygen vacancies, leading to a structure as $Pb_8NbTi_7O_{23}$ ($PbNb_{0.125}Ti_{0.875}O_{2.875}$, $PNTO_{2.875}$). DFT calculations were performed within the generalized gradient approximation (GGA) with the Perdew-Burke-Ernzerhof (PBE) exchange-correlation functional[33] as implemented in the VASP[34]. The projector augmented wave method (PAW)[35] was used with the following electronic configurations: $5d^{10}6s^26p^2$ (Pb), $3s^23p^63d^24s^2$ (Ti), $4s^24p^64d^45s^1$ (Nb) and $2s^22p^4$ (O). An effective Hubbard term $U_{eff}$ = U–J using Dudarev's approach[36] with $U_{eff}$ = 3.27 eV was included to treat the Ti 3d orbital, and a 520 eV cutoff energy of the plane-wave basis set was used for all calculations. For structure optimizations, atomic positions were considered relaxed for energy differences up to $1 \times 10^{-6}$ eV and all forces were smaller than $1 \times 10^{-2}$ eV Å$^{-1}$. A $9 \times 9 \times 9$ gamma-centered $k$-point mesh was used, and denser $k$-point mesh was used for density of states calculations. The atomic structure was visualized using the VESTA package[37].

## 4. RESULTS AND DISCUSSION

**4.1. XRD and RSM.** The X-ray diffraction (XRD) $\theta$−$2\theta$ scan results for PNTO$_{3-\delta}$ films grown in 6 Pa and 4 Pa oxygen partial pressures suggest the good qualities of these films without any secondary phase (Supporting Information, Figure S1a). The $Q_x$ values of the spots of the films and substrates in the X-ray diffractometry reciprocal space mapping (RSM) are almost the same (Supporting Information, Figure S1b), indicating coherent growth of these films. The single spot at (103) diffraction implies no phase separation in the films. The STEM results (Supporting Information, Figure S2a) present a continuous and sharp interface, implying the epitaxial deposition of these PNTO$_{3-\delta}$ films. Moreover, the larger lattice parameter $c$ to $a$ in the STEM image (Supporting Information, Figure S2b) verifies the tetragonal structure of PNTO$_{3-\delta}$ films deduced from Figure S1. Electron diffraction result verifies that the PNTO$_{3-\delta}$ films are monocrystalline (Supporting Information, Figure S2c).

**4.2. Ferromagnetism.** The magnetic properties of these PNTO$_{3-\delta}$ films were examined using a PPMS in the temperature range of 10 K to 300 K. Figures 1a and 1c illustrate the in-plane (IP) magnetic-field-dependent magnetization (*M*) for PNTO$_{3-\delta}$ films fabricated in the oxygen pressures of 6 Pa and 4 Pa, respectively. Different from the films deposited under 8 Pa oxygen pressure, in which only diamagnetic component were observed with the paramagnetic (PM) nature of the films[38], the PNTO$_{3-\delta}$ films become ferromagnetic (FM) when the deposition oxygen pressure was reduced to 6 Pa, with saturation magnetizations ($M_s$) of ~2.7 emu/cm$^3$ at 300 K and ~5.4 emu/cm$^3$ at 10 K. As the deposition oxygen pressure further decreased to 4 Pa, the ferromagnetic properties of these films get ameliorated, exhibiting an improved squareness in magnetic hysteresis loops and enhanced $M_s$ of ~6.3 emu/cm$^3$ at 300 K and 7.9 emu/cm$^3$ at 10 K.

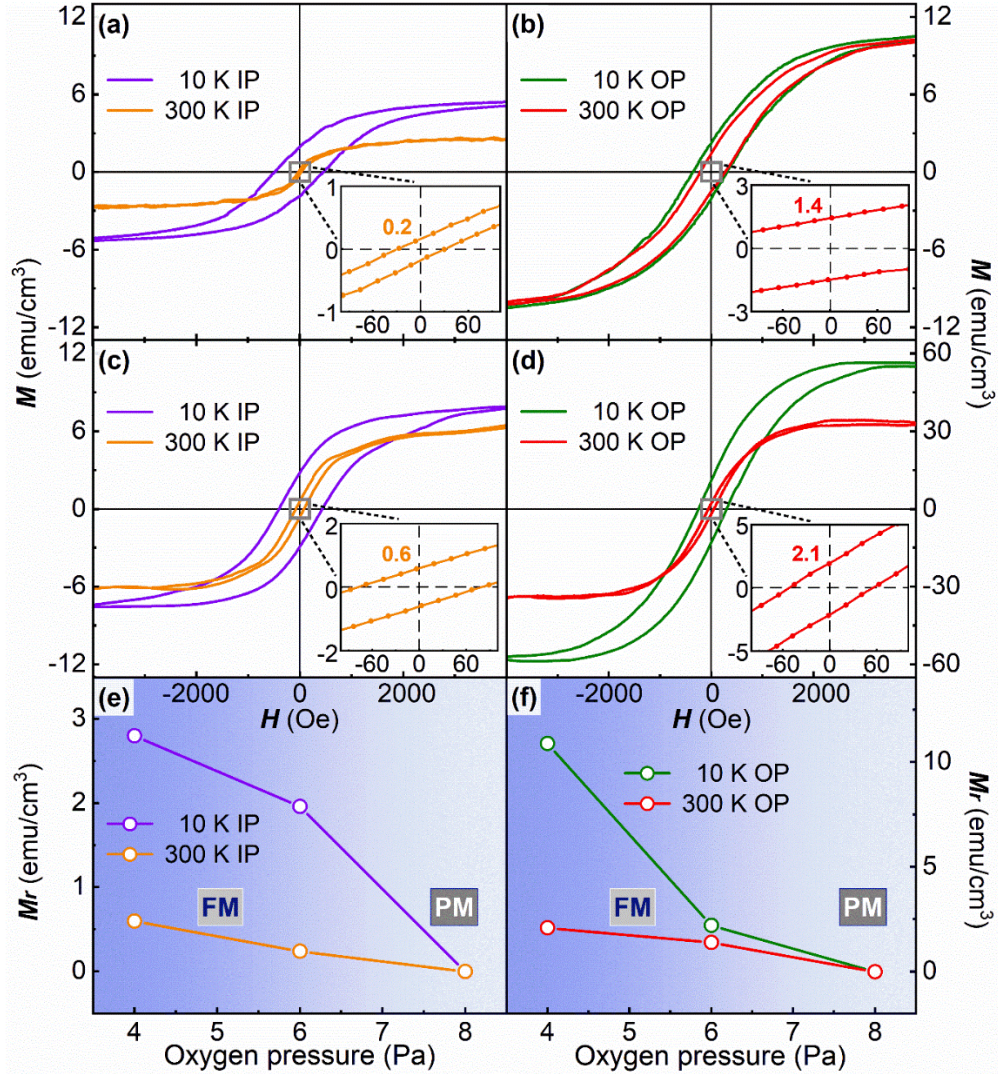

**Figure 1.** Magnetic properties of PbNb$_{0.12}$Ti$_{0.88}$O$_{3-\delta}$ (PNTO$_{3-\delta}$) films. The in-plane (IP) magnetic-field-dependent magnetization ($M$) measured in 300 K and 10 K for PNTO$_{3-\delta}$ films deposited in (a) 6 Pa and (c) 4 Pa. (b) and (d) The corresponding out-of-plane (OP) magnetic hysteresis loops for the films in (a) and (c). The insets in (a-d) show the zoom-in images of the magnetic hysteresis loops captured in 300 K marked by the grey squares. In the insets, the dots are experimental data and the solid lines are for eye guide. The numbers denote the remnant magnetization ($M_r$). (e) IP and (f) OP $M_r$ dependences on the deposition oxygen pressures.

The corresponding OP magnetic properties are displayed in Figures 1b and 1d. Similarly, ferromagnetic properties cannot be found in the films grown under 8 Pa[38] but in 6 Pa and 4 Pa. Meanwhile, saturation magnetizations rise from 10.3 emu/cm$^3$ to 33.6 emu/cm$^3$ at 300 K, from 10.8 emu/cm$^3$ to 56.3 emu/cm$^3$ at 10 K with the deposition oxygen pressure dropping from 6 Pa to 4 Pa. Figures 1e and 1f show the remnant magnetization ($M_r$) dependences on the deposition oxygen pressures in IP and OP directions, respectively, collected from Figures 1a-1d and our previous work[38]. As the deposition pressure changes from 8 Pa to 6 Pa and 4 Pa, the $M_r$ enlarges with a transition of PNTO$_{3-\delta}$ films from PM to FM. These results prove that the PNTO$_{3-\delta}$ films synthesized under 6 Pa and 4 Pa oxygen pressures exhibit distinct ferromagnetisms, even at 300 K, indicating the achievement of room-temperature ferromagnetisms. Meanwhile, these PNTO$_{3-\delta}$ films manifest asymmetric magnetic characteristics with the OP $M_s$ and $M_r$ bigger than those alone IP. The ferromagnetisms of these PNTO$_{3-\delta}$ films maintain as the temperature raised up to 390 K.

**4.3. Ferroelectricity and Metallic behavior.** The polar structure of PNTO$_{3-\delta}$ films fabricated under 6 Pa and 4 Pa oxygen pressures were investigated using annular bright-field (ABF) STEM, shown in Figures 2a and 2d, respectively. The reconstructed images both reveal the downward displacements of oxygen atoms from the face center of Pb along (001) direction (red arrows) and the slightly shifting of Nb$_{0.12}$Ti$_{0.88}$ atoms upward from center of the oxygen octahedron (olive arrows), forming polar displacements along *c* direction in these domains, illustrated by the transparent arrows. The switchable polarizations of both samples were further examined using PFM, exhibited in Figures 2b and 2e. Sharp contrasts could be easily distinguished in the PFM phase images attained from the surfaces of these PNTO$_{3-\delta}$ films within an area of 2.6 × 2.6 μm$^2$.

Second checks of the switched status were conducted after 30 min, and the good contrasts remained, implying the stable switchable polarizations in these films. The SHG technology was also utilized to investigate the noncentrosymmetric structures of these films, shown in Figures 2c and 2f. The SHG anisotropic patterns were checked in *p*- and *s*-out configurations[32]. Theoretical fittings of the results confirm that the point groups of both PNTO$_{3-\delta}$ films are *4mm*, in accordance with the tetragonal structures attested from XRD, RSM, and STEM analyses. The corresponding macroscopic ferroelectricities of these samples were studied using a ferroelectric tester, shown in Figures 2g and 2h. Although the remnant polarization values obtained in Figures 2g and 2h may not be accurate due to the leakage of the current caused by the free carriers, these results clearly show that the heavily doped PNTO$_{3-\delta}$ films have nonzero remnant polarizations when the electric field drops to zero. This demonstrates that, more than the existence of internal polar distortions in the heavily doped PNTO$_{3-\delta}$ films, the polar distortions are not completely screened out by the itinerant carriers. The results above verify the room-temperature polar structure and ferroelectricity of these PNTO$_{3-\delta}$ films.

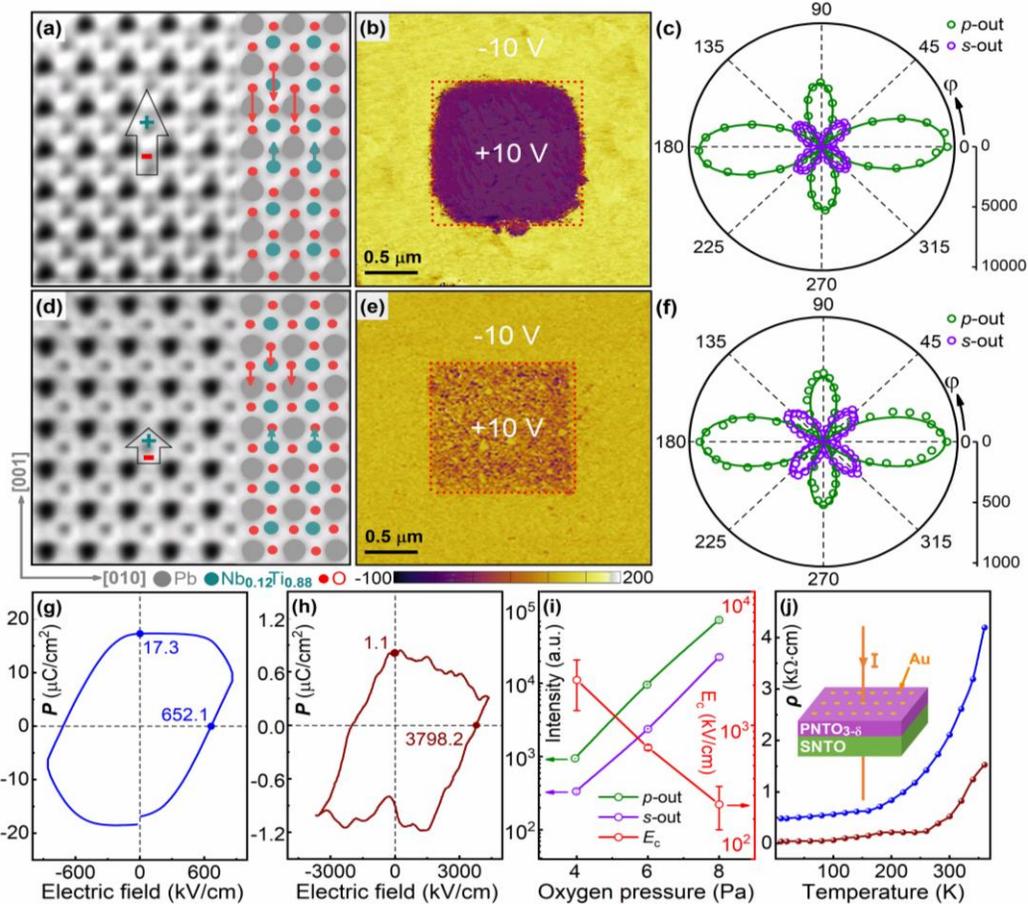

**Figure 2.** Polar displacements, ferroelectricities, and electric transport properties of PbNb$_{0.12}$Ti$_{0.88}$O$_{3-\delta}$ (PNTO$_{3-\delta}$) films. Annular bright-field scanning transmission electron microscopy (ABF-STEM) images of the PNTO$_{3-\delta}$ films grown in (a) 6 Pa and (d) 4 Pa. The transparent arrows demonstrate the polarization directions in these domains. The lengths of these red, olive, and transparent arrows suggest the relative magnitude of the O-Pb, Nb$_{0.12}$Ti$_{0.88}$-O polar displacements, and the electric dipole moments. Piezoresponse force microscopy (PFM) phase images gained from PNTO$_{3-\delta}$ films fabricated in (b) 6 Pa and (e) 4 Pa. Double squares denote the switched areas by a conductive tip with ±10V, causing a polarization pointing downward for the central square (1.3 × 1.3 μm$^2$) and upward for the outer square (2.6 × 2.6 μm$^2$). Second-harmonic generation (SHG) $p$- (olive) and $s$-out (violet) polar plots for PNTO$_{3-\delta}$ films grown in (c) 6 Pa and (f) 4 Pa. $\varphi$ is the polarization angle of the incident laser. The circles are experimental data and the solid lines are theoretical fittings by *4mm* point group. Electric-field-dependent polarizations (***P***) of the PNTO$_{3-\delta}$ films synthesized in (g) 6 Pa and (h) 4 Pa. (i) SHG peak intensities and ferroelectric coercive field (***E**$_c$*) variations as a function of deposition oxygen pressures. The numbers indicate the remnant polarization and ***E**$_c$*. (j) The temperature-dependent resistivities of the PNTO$_{3-\delta}$ films grown in 6 Pa (blue) and 4 Pa (red) oxygen pressures. The spheres are experimental data and the solid lines are for eye guide. The inset in (j) is the schematic illustration of measurement configuration.

Figure 2i displays the SHG peak intensities and ferroelectric coercive fields ($E_c$) dependences on deposition oxygen pressures, deduced from Figures 2c, 2f, 2g, 2h, and our previous work[38]. As the deposition partial pressure reduces from 8 Pa to 6 Pa and 4 Pa, the SHG intensity drops two orders of magnitude, illustrating the attenuation of noncentrosymmetric polar structure of these films. Besides, the increase of $E_c$ reveals that it is more arduous to reverse the polarization in these films deposited in lower oxygen pressures. These results agree well with the corresponding strength of the polar displacements in STEM, phase contrasts in PFM, and remnant polarizations in ferroelectricity examinations above, which all indicate larger electric polarization in the films deposited under 6 Pa oxygen pressure than those under 4 Pa. The electric transport properties of these PNTO$_{3-\delta}$ films were measured using PPMS. Figure 2j shows the temperature-dependent resistivity ($\rho$) from 10 K to 360 K. The resistivity of both PNTO$_{3-\delta}$ films increases as the temperature rising, yielding the metallic behaviors. Moreover, the films deposited under 4 Pa oxygen pressure exhibit a better conductivity than those under 6 Pa.

**4.4. Oxygen Vacancies.** Along with the decrease of oxygen partial pressure during the deposition, PNTO$_{3-\delta}$ films changed into FM from PM[38] with the strengthening of *$M_s$* and *$M_r$*. Meanwhile, their conductivity improves, causing the non-neglectful leaky currents while switching the polarization of these low-pressure-synthesized samples and the weakening of noncentrosymmetric polar structure. These tendencies are attributed to the increase of $V_o$[39-46] which were ineluctably induced into the films when depositing in low oxygen pressures. The delocalized electrons induced by the oxygen deficiencies should contribute to the conductivity and screen the polar displacements.

In order to know the oxygen content in these PNTO$_{3-\delta}$ films, the ABF-STEM technology was engaged. The O-Nb$_{0.12}$Ti$_{0.88}$-O chains were chosen at the same depth from the interface to get rid of the influence of layer-dependent oxygen content variation[47,48]. Figures 3a, 3c, and 3e display the cross-sectional ABF-STEM micrographs of an area ~6 × 6 nm$^2$ of films synthesized under 8 Pa, 6 Pa, and 4 Pa oxygen pressures, respectively. The oxygen concentrations are correspondingly presented in the line profiles of the O-Nb$_{0.12}$Ti$_{0.88}$-O chains in Figures 3b, 3d, and 3f, respectively. Since the oxygen concentration is proportional to the depth of the valleys[47-49] pointed out by red arrows, the line profiles in Figures 3d and 3f clearly suggest a deficiency of oxygen content and considerable $V_o$ in PNTO$_{3-\delta}$ films deposited in 6 Pa and 4 Pa oxygen pressure, compared with those in 8 Pa which is abundant with oxygen[38].

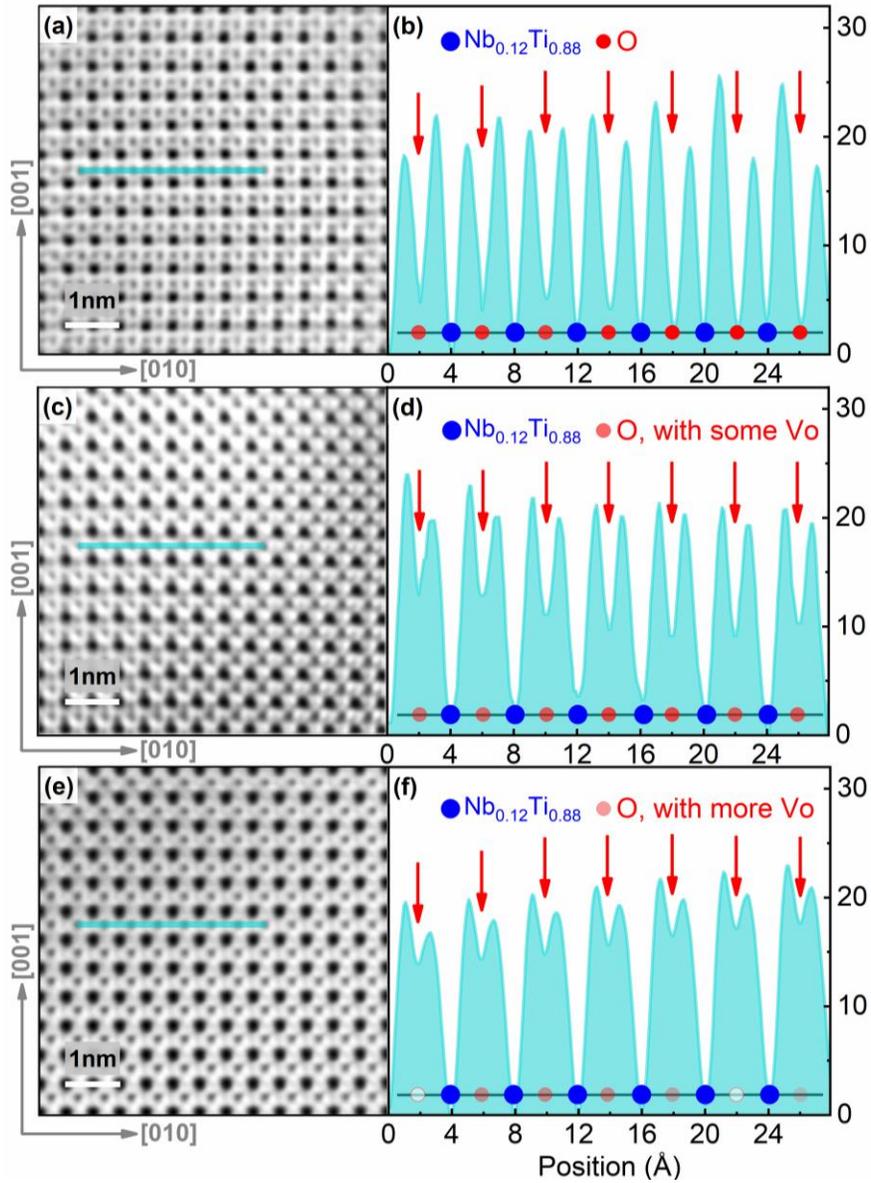

**Figure 3.** Oxygen concentration ins PbNb$_{0.12}$Ti$_{0.88}$O$_{3-\delta}$ (PNTO$_{3-\delta}$) films deposited in different oxygen pressures. The cross-sectional ABF-STEM images of PNTO$_{3-\delta}$ films grown in (a) 8 Pa, (c) 6 Pa, and (e) 4 Pa oxygen pressures. (b), (d), and (f)The corresponding line profiles of O-Nb$_{0.12}$Ti$_{0.88}$-O chains marked in (a), (c), and (e) by solid cyan lines. The blue and red dots represent the Nb$_{0.12}$Ti$_{0.88}$ and oxygen atoms, respectively. The different transparency of the red color of oxygen atoms in (b), (d), and (f) denote none, some and more oxygen vacancies (V$_o$). All data have been normalized to the corresponding Pb atoms mean values.

**4.5. First-principles Calculation.** First-principles calculation was performed to inspect the coexistence of ferroelectricity, ferromagnetism, and metallic behavior in the Nb-doped PbTiO$_3$ films with oxygen deficiency. Figure 4a shows the relaxed structure of the supercell. The red arrows represent the polar displacements of oxygen atoms shifting downward from Pb face center, while the olive arrows illustrate the fractional polar displacement of Nb atoms relative to the oxygen octahedral center. These polar displacements are accordant with what were observed in STEM imaging. The yellow areas present the anisotropic distribution of free electrons, with a charge density isovalue of 0.004 eÅ$^{-3}$, from which we can see the free electrons mainly locate at the $d_{xy}$ orbitals of Nb and Ti atoms. The polar displacements of Pb and O in PNTO$_{3-\delta}$ films are not extinguished because the Fermi surface originates from Ti (3$d$ t$_{2g}^1$ configuration) and Nb (4$d$ t$_{2g}^1$ configuration) orbitals while Nb/Ti exhibit negligible contributions to the polarization[27,50,51], similar to the case in polar metal Ca$_3$Ru$_2$O$_7$[8]. Figure 4b shows the spin-resolved total density of states (TDOS) of PNTO$_{2.875}$, in which the Fermi level intersects the conductive band, verifying the metallicity. Compared with PbTiO$_3$[52] whose Fermi level locates at the conductive band minimum, the sources of the free electrons in PNTO$_{3-\delta}$ films are both doped Nb and oxygen vacancies. Moreover, the Fermi level passes through the spin-up channel only, resulting in a type-IA half metal[53] for 12.5% Nb-doped PbTiO$_3$ with V$_o$. The characteristic that the low resistivity is only presented for spin-up electrons, as the exfoliated MnPSe$_3$ nanosheet with carrier doping[54], would offer a new design for spintronic devices.

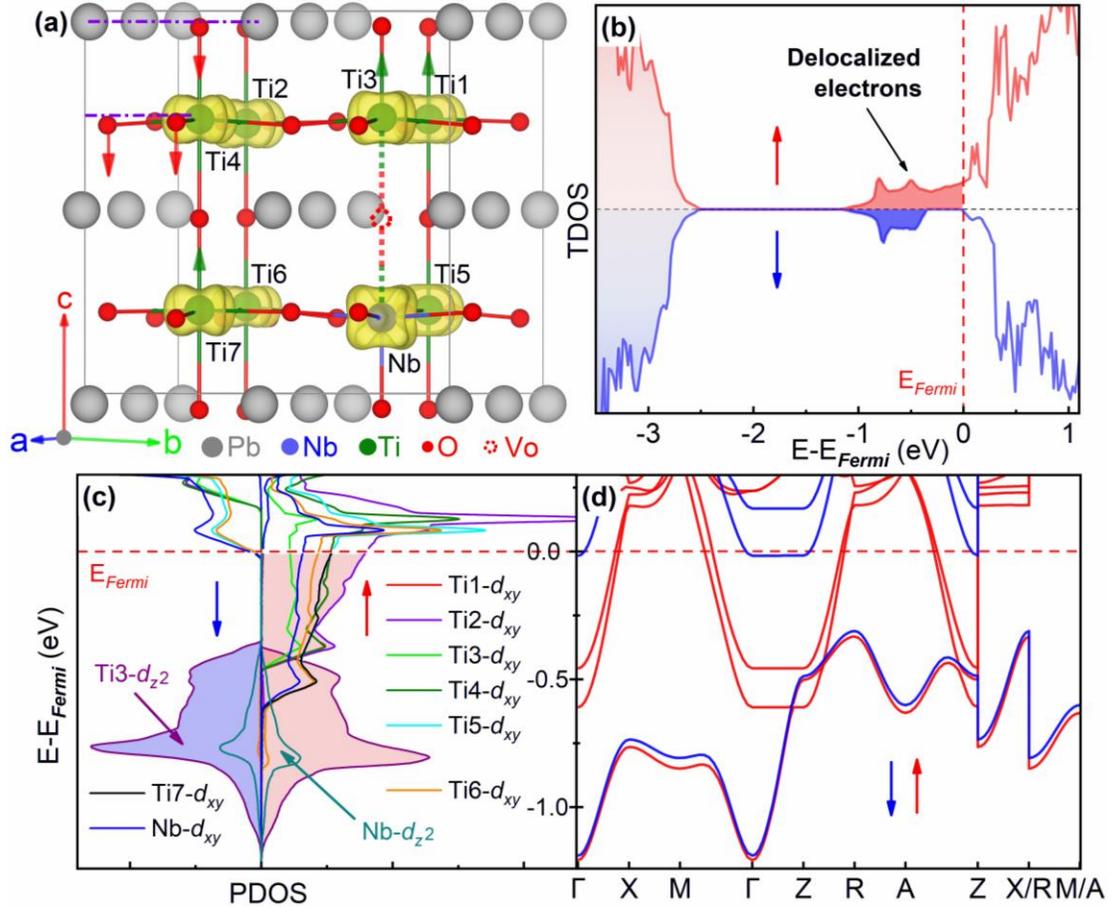

**Figure 4.** First-principles calculation results for PbNb$_{0.125}$Ti$_{0.875}$O$_{2.875}$ (PNTO$_{2.875}$) with oxygen vacancies (V$_o$). (a) Stable structure of PNTO$_{2.875}$. The grey, blue, olive, and red dots denote the Pb, Nb, Ti, and O atoms, respectively. The red dotted circle represents V$_o$. The violet dashed lines imply the vertical positions of Pb face center along the *c* axis. The red and olive arrows indicate the polar displacements of O-Pb and Ti-O, respectively. The yellow areas represent the free electron distribution in this primitive cell. (b) Spin-resolved total density of states (TDOS). The delocalized electrons are presented as the colorfully filled area, see the black arrow. (c) The atomic-projected density of states (PDOS) within an energy interval from -1.2 eV to 0.3 eV. DOS for other atoms and orbitals are neglectable within this energy region. (d) The corresponding band structure to (c). The red and blue arrows in (b), (c), and (d) indicate the spin up and spin down states, respectively. The dashed lines represent the Fermi level.

We believe the polar survives due to the weak coupling between the conductive electrons/holes at the Fermi level and soft mode phonons[52,55,56], so that the conductive electrons/holes are not able to screen the polar completely, which makes the polar switchable in the films.

Figure 4c displays the atomic-projected density of states (PDOS) near the Fermi level, revealing that the free electrons locate at Nb and Ti atoms. It further unveils that the free electrons with lower energy mainly originates from the $d_z^2$ orbitals of Ti3 and Nb atoms. The $d_z^2$ orbitals of Ti3 and Nb atoms manifest a lower energy than their $d_{xy}$ orbitals since the oxygen atom was removed from the site between them to simulate the V$_o$, as the dotted circle indicates. The calculation for other cases in which oxygen atom was removed from different positions were also performed and the results are similar. Those free electrons with higher energy distribute at the $d_{xy}$ orbitals of Nb and Ti atoms, the same as the occasions in Nb:BaTiO$_3$[11] and Nb:PbTiO$_3$[9], because the Jahn-Teller effect and electrostatic field in each Ti-O and Nb-O octahedral complex make the spin-up state of the $d_{xy}$ orbitals of Ti/Nb atoms presenting the second lowest energy except the $d_z^2$ orbitals of Ti3 and Nb atoms, leading to the ferromagnetism in the PNTO$_{3-\delta}$ films. Figure 4d shows the band structure of PNTO$_{2.875}$ corresponding to Figure 4c, in which the Fermi level intersects spin-up band (red) only, verifying the half metallicity in 12.5% Nb-doped PbTiO$_3$ films with V$_o$.

## 5. CONCLUSION

In conclusion, we achieved the room-temperature coexistence of ferromagnetism, ferroelectricity, and metallic behavior in PNTO$_{3-\delta}$ films by varying the deposition oxygen partial pressure. For these films grown in 6 Pa and 4 Pa oxygen pressures, the magnetic hysteresis loops are overt, accompanied with distinct metallic behaviors. Meanwhile, we manifest a tetragonal structure with the switchable polar distortions not intensively screened. This simultaneous appearance of multiferroicity and metallic behavior is also verified by first-principles calculation. The determination of relaxed structure of 12.5% Nb-doped PbTiO$_3$ with oxygen vacancies shows apparent polar displacements. The asymmetric non-zero spin-resolved density of states at the Fermi level reveals the existence of itinerant electrons and spin polarization in the PNTO$_{3-\delta}$ films. Additionally, the half metallicity of PNTO$_{3-\delta}$ films exhibits potential diverse electric properties for electrons with different spin orientations. These findings offer a new aspect to explore and design spintronic multiferroic devices.

**Supporting Information**.

Structural characterization of the heterostructure, cross-sectional STEM diagram of the interface, and the electron diffraction pattern.

AUTHOR INFORMATION

**Corresponding Author**

*E-mail: kjjin@iphy.ac.cn (Kuijuan Jin)

**Author Contributions**

The manuscript was written through contributions of all authors. All authors have given approval to the final version of the manuscript. ‡These authors contributed equally: H. B. Y[‡]. and J. S. W[‡]. K. J. J. conceived the project. H. B. Y. performed the sample characterization, SHG analysis, electric and magnetic properties measurements. J. S. W. carried out the ferroelectricity test and PFM imaging. Q. H. Z. and G. L. did the STEM experiments. H. B. Y., J. S. W., and V. P. prepared the $PNTO_{3-\delta}$ films. W. N. R. conducted the first-principles calculations. K. J. J. supervised the experimental and theoretical studies. All authors discussed the results and commented on the manuscript.

**Notes**

The authors declare no competing financial interest.

ACKNOWLEDGMENT

This work was supported by the National Key Basic Research Program of China (Grant No. 2017YFA0303604), the National Natural Science Foundation of China (Grants Nos. 11721404, 51761145104, and 11674385), the Key Research Program of Frontier Sciences of the Chinese Academy of Sciences (Grant No. QYZDJ-SSW-SLH020), and Postdoctoral Science Foundation of China (Project No. 2019M650877). We acknowledge the inspiring talks with Dr. Chao Ma from Shandong Agricultural University for first-principles calculation.

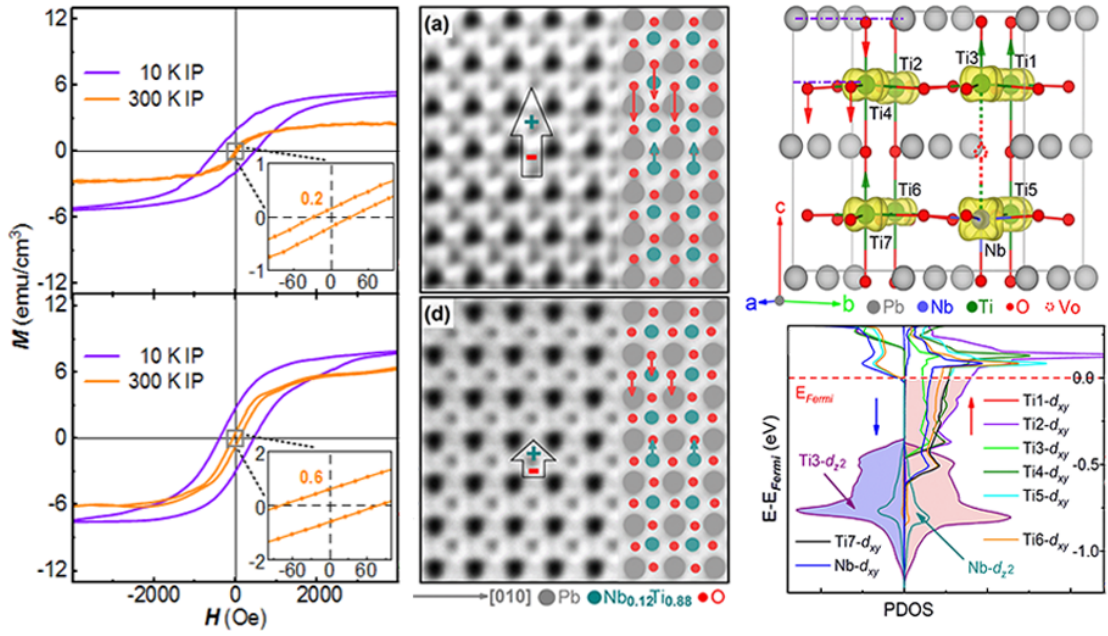

TOC Graphic